# A NEW APPROACH FOR FORMAL BEHAVIORAL MODELING OF PROTECTION SERVICES IN ANTIVIRUS SYSTEMS

Monire Norouzi[1], Saeed Parsa[2*] and Ali Mahjur[3]

[1]Department of Computer Engineering, Shabestar Branch, Islamic Azad University, Shabestar, Iran
[2*]Department of Computer Engineering, Iran University of Science and Technology(IUST), Tehran, Iran
[3]Malek Ashtar University of Technology, Tehran, Iran

## ABSTRACT

*Formal method techniques provides a suitable platform for the software development in software systems. Formal methods and formal verification is necessary to prove the correctness and improve performance of software systems in various levels of design and implementation, too. Security Discussion is an important issue in computer systems. Since the antivirus applications have very important role in computer systems security, verifying these applications is very essential and necessary. In this paper, we present four new approaches for antivirus system behavior and a behavioral model of protection services in the antivirus system is proposed. We divided the behavioral model in to preventive behavior and control behavior and then we formal these behaviors. Finally by using some definitions we explain the way these behaviors are mapped on each other by using our new approaches.*

## KEYWORDS

*Antivirus system, protection services, behavioral modeling, formal method;*

## 1. INTRODUCTION

Today, antivirus software is very important in business and software development, because any computer the system should install a security application on itself to maintain regularize and security of its data [1]. In the recent years, many attacks are occurred to home the systems, bank servers and military systems by Viruses and Malwares [2]. Information maintenance and prevent to data accessibility is the cause of attacking and destroying data which has been occurred by invasive malware widely and suddenly [3]. So, users need to powerful security applications which secure their systems against the attack of viruses and malwares [4].

The attacks in the system security section is divided into two parts [5]. The first set of attacks are enabled to eliminate security files or security applications of a computer the system. After disabling these files and applications, these attacks access to important data in the computer system. The second set of attacks can remove a set of special files and destroying important files. These attacks are executed and handled from intelligence agencies and hacker on computer systems of the users. These attacks have not needed to destroy the system security of computer. But, these attacks are executed and run by hiding themselves in URL web addresses or in created files using routing software such as Microsoft office, adobe reader, win zip, etc. By notice to these attacks, a computer system need to an online security system. Of course, an offline security system prevent to data accessibility in computer system slightly. Unfortunately, by progression of technology and software system complexity attackers have discovered some new vulnerable areas in the systems that informally called "holes". So, by this ways they disturb the system security.
By the above reasons, we find that testing and verifying the security applications such as the antivirus systems is very important and essential in security discussion of computer systems [6]. There are some antivirus applications in software development market such as Bitdefender, Kaspersky, Avira and etc. that each application try to compete with each other by presenting more services and easy updating. Of course, computer viruses [7], Spywares [8], Trojans, Worms [9] and other new malwares debut every day.

In this paper, a protection services in Antivirus System approach has been presented. In particular, we separate protection services in Antivirus System into two types: preventive behavior and control behavior based on behavioral modeling methods of Hansen, Virtanen et al. (2003). The interactions between these two behaviors are modeled as characteristics process. The process role is to keep the couple behaviors synchronized. By analyzing logical problems and checking behavior specifications, it is possible to verify the protection services in Antivirus System approach. In particular, the contributions of this paper are:

- Proposing an ideal Antivirus System based on Avira Antivirus approach.
- Presenting an Antivirus System behavior model to couple preventive and control behaviors of Antivirus System approach.
- Facilitating the mapping process between these two behaviors by means of the formal verification approach based on Binary Decision Diagram (BDD) (Clarke, Grumberg et al. 1999).

The rest of this paper is organized as follows. Section 2 reviews the related works and correlated studies in formal verification and Antivirus System approach. Section 3 describes theformalizing preventive and control behaviors of the Antivirus System as well as defining some essential concepts and notations to formalize these behaviors. Finally, conclusions and future works are provided in Section 4.

## 2. RELATED WORKS

Defining an anti-virus model is used to designing antivirus applications in software development, so it is not a new topic. But, formalizing and verifying an antivirus model by using formal verification methods is a new idea in software development. Some researchers used to formal verification methods for analysis and verification of software systems. For example:

C. Livadas and et al investigate how formal techniques can be used for the verifying hybrid systems. Then by presenting the hybrid I/O automaton model they applied to the specification and verification of hybrid systems [10].

Qianchuan Zhao and et al in their research show how CTL specifications for a statechart can be verified using a finite-state model checker. In this paper, authors use to kripke structure that provides the formal relationship between the proposed model and the statechart structure and CTL formulas for showing formal semantics in CTL specification [11].

A new model has been presented for verifying a website by using LTL logic and Formal method techniques. The authors model web pages of a website as states and convert HTTP protocols of the website to transition between states. This paper shows useful relationships between a proposed model and kripke structure. Of course, all of these papers have same procedure for formalizing their models and creating kripke structures. But in our paper, we describe how expected specifications of system verify by using defined kripke structure and proposed relations between proposed model and CTL logic [12].

J.A. Morales et al present a formal model of virus transformation that enables variation traceability by using four antivirus solutions for handheld devices. Also, they test their formal model for antivirus software that some viruses attack to this software. In this paper, formalizing antivirus model use to testing some solutions against virus attacks [13].

Andronick, J. et al considered a new approach to verification of a smart card embedded operating system. They proved a C source program against supplementary annotations and generated a high-level formal model of the annotated C program that was used to verify certain global security properties. This paper is focused on modeling smart card security in Embedded Source Codes [14].

Heitmeyer, C.L., et al presented verifying a system's high-level security properties. Their approach is focused on computer security by using antivirus systems rather than security properties of software systems Heitmeyer, et al. [15].

Yeung, W.L. presented a formal verification approach to describe the behavioral problems such as deadlocks and live locks in multi-agent systems. They proposed a model of the contract negotiation process in multi-agent systems. The proposed model translated to XML Metadata Interchange (XMI) documents and converted to Communicating Sequential Processes (CSP) scripts by using ArgoUML tool. Also the Failures-Divergences Refinement (FDR) tool have been used to implement the model [16].

Yeung, W.L. described a formal and visual modeling approach for choreography-based web services composition and conformance verification. Apart from Web Services Choreography Description Language (WS-CDL) and Web Services Business Process Execution Language (WS-BPEL), their approach also supported the use of visual modeling notations such as UML in modeling choreographies and orchestrations. The architecture of proposed web services is presented based on Simple Object Access Protocol (SOAP) and Web Services Description Language (WSDL). Then a choreography model of the contract negotiation process is presented by using Unified Modeling Language (UML) tool. Also they translated proposed model in UML to CSP by using XMI transformation. After translating, the mathematical approaches executed on translated model and the model is verified by using FDR model checker. Finally, integration of semantic web service technologies into the framework is also being considered [17].

As another research in this scope Bentahar, J., et al modeled the composite web services based on a division of interests between operational and control behaviors. Some favorable properties such as deadlock freedom, safety and reachability has been analyzed. The proposed behaviors have been converted to Kripke structure by using model checking techniques based on BDD. Also the Kripke structure models have been translated to SMV code by means of Java converter tool. Then the models were verified by using NuSMV model checker [18].

Souri et al proposed an adapted resource discovery approach to address multi-attribute queries in grid computing. They presented a behavioral model for their proposed approach that separate into data gathering, discovery and control behaviors. So, they used to kripke structure for modeling these behaviors and verify their behavioral models by using NuSMV model checker[19].

Shukla, J.B., et al proposed a nonlinear mathematical model and analyzed to study the role of antivirus program to clean an infected computer network. Their model has been proposed by employing the concept of epidemics, where nodes in the network are considered as populations.

International Journal in Foundations of Computer Science & Technology (IJFCST), Vol.4, No.3, May 2014

The total numbers of nodes have been separated into vulnerable nodes, infected nodes, and protected nodes. The variable representing the number of antivirus programs considered in the model has been assumed to be proportional to the number of infected nodes. Their model has simulated by MAPLE and MATLAB software and there is no any scientific reason for modeling spread of viruses and infected files in computer networks [20].

## 3. FORMAL BEHAVIORAL MODELLING

In this paper, a protection services in Antivirus System approaches has been presented. In particular, we separate protection services in Antivirus System into two types: preventive behaviorand control behaviorbased on behavioral modeling methods of Hansen, et al. [21]. The interactions between these two behaviors are modeled as mapping process. The process role is to keep couplebehaviors synchronized. In particular, the contributions of this paper are:

- Proposing an ideal Antivirus System based on Avira Antivirus approaches.
- Presenting an Antivirus System behavior model to preventive and control.
- Presenting a mapping process between the preventive and control behaviors.
- Formalizing preventive and control behaviors.

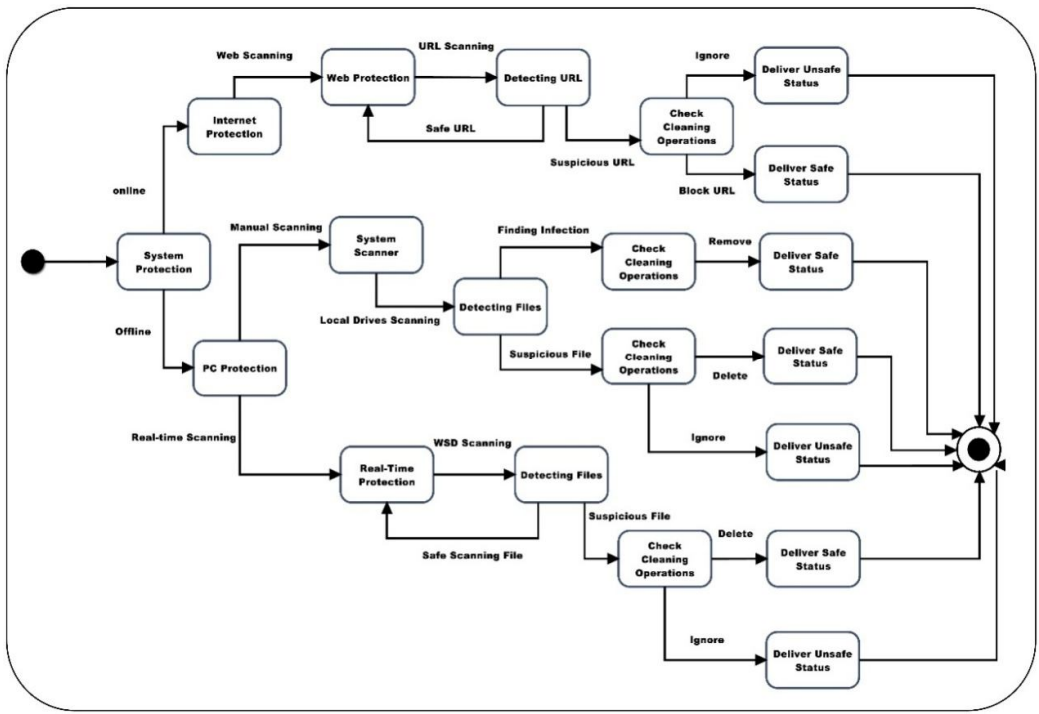

Figure 1. Preventive behavior of the antivirus system

Figure 1 describes preventive behavior of AVIRA antivirus software in a simple statechart. The system utilizes security rules of the antivirus system to establishing four specific approaches that include *Protection*, *Detection*, *Identification* and *Removal*. The preventive behavior starts protection approach by initial state system protection and specifies system mode by checking online or offline mode. The offline mode shows PC Protection and includes protection approach for two states Real-time Protection and System Scanner. In the state Real-time Protection, the system executes protection operations automatically in the windows system directory (WSD) section because the important systematic files hold this path:" C:\windows\system32".

The state detecting file as a type of Detection approach discover any type of infection in suspicious file and send it to state check cleaning operation as a type Identification approach. If the infected file deletes then the Removal approach is executed and system receives safe result in this process. If suspicious file ignore for any reason then system receive unsafe result in this process. The next section of offline mode is manual scanning that the state system scanner executes protection process on local drives, removable drives and local hard discs by state detecting file as a type of Detection approach.

The recycling operation divides to two parts. In the first part, by finding infection position the system performs Identification approach and removes the infection from file so the file is safe and Removal approach is executed. In the second part, a suspicious file is found and state check cleaning operations executes Identification approach. If this file is deleted then Removal approach receives safe result otherwise the system receives unsafe result.

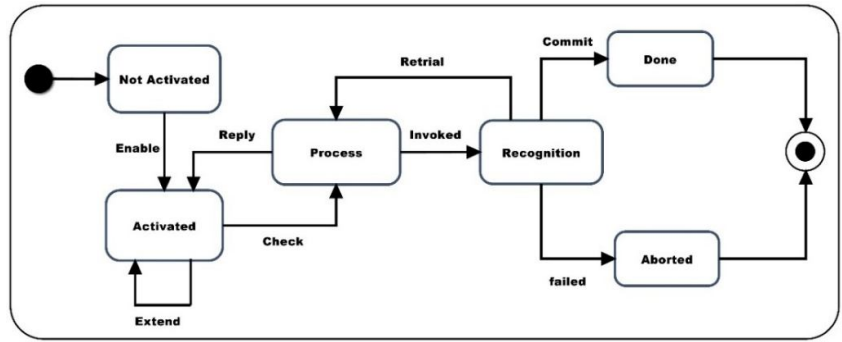

Figure 2. Control behavior of the antivirus system

The control behavior navigates the execution flow of approaches of antivirus system. In the figure 2, the control behavior of the antivirus system presents a number of states extracted from preventive behavior by using the Buchi automata. These states include *Not Activated*, *Activated*, *Process*, *and Recognition*, *Done* and *Aborted*.First, by activating antivirus system the control behavior will be enabled and the all of protections like system protection, internet protection, web protection, PC protection and real-time protection are enabled. URL detecting and file detecting will be checked and by finding suspicious files or suspicious URL or infected file they will be removed or ignored in Recognition state. The state Done occurs if the action in Recognition state deliver safe state by removing the infected or suspicious file and the state Aborted occurs when the action in Recognition state is ignored the infected or suspicious file.

**Definition 1** (path in antivirus system behaviors): A path $p_{i \rightarrow j}$ in an antivirus system behaviors is a finite sequence of the states and transitions starting from state $s_i$ and finishing at state $s_j$, denoted as:

$$p_{i \rightarrow j} = s_i \xrightarrow{l_i} s_{i+1} \xrightarrow{l_{i+1}} s_{i+2} \ldots s_{j-1} \xrightarrow{l_{j-1}} s_j \text{ such that } \forall\ m \in \{i, j-1\}: (s^m, l^m, s^{m+1}) \in T.$$

For example, in figure 2, *Not Activated* $\xrightarrow{l_1}$ *Activate* $\xrightarrow{l_2}$ *Process* $\xrightarrow{l_3}$ *Recognition* $\rightarrow$ $\xrightarrow{l_4}$ *Aborted* $\xrightarrow{l_5}$ *End* is a path in the control behavior of the antivirus system.

This section present an antivirus behavior model using formal and model checking techniques, which guides antivirus approaches for any Coordination scenario and separates its behavior into preventive behavior and control behavior. We use statechart semantics to modeling preventive

81

and control behaviors. Of course other formalism approaches can be used such as Petri nets techniques [Robert G. Pettit, Xitong Li].

Now, we illustrate a formal description of the antivirus behavioral model. The preventive behaviordemonstrates the security rules that implement specific approaches for functioning of an antivirus system. Also the control behavior navigates the execution flow of security rules of an antivirus system. We define these behaviors as follow:

**Definition 2** (antivirus system behavior): The antivirus system behavioris a 4-tuple AB= (S, $s_0$, L, R) where:

- S is a finite set of states.
- $s_0 \in$ S is the initial state.
- L is a set of transition labels.
- T $\in$ S × L × S is the transition relation. For showing a transition relation we have t = ($s_i$, l, $s_j$)

that t $\in$ T, $s_i$ and $s_j \in$ S and l $\in$ L. This transition relation describe existing a relation between $s_i$ and $s_j$ by label l.

The preventive and control behaviors of an antivirus system have retrieved from the antivirus system behavior.

**Definition 3** (preventive behavior): The preventive behavioris a 4-tuple $Pr_B$ = ($S_{pr}$, $s_{pr}$, $E_{pr}$, $A_{pr}$) where:
- $S_{pr}$ is a finite set of preventive behaviorstates.
- $s_{pr} \in S_{pr}$ is the initial state.
- $A_{pr}$ is a specific approaches of the antivirus behavior.
- $E_{pr}$ is a finite set of preventive behavior events.
- The approach $A_{pr}$ is demonstrated by $s_{pr} \xrightarrow{e} s'_{pr}$: $a_{pr}$

**Definition 4** (control behavior): The control behavioris a 4-tuple $Co_B$ = ($S_{co}$, $s_{co}$, $E_{co}$, $A_{co}$) where:
- $S_{co}$ is a finite set of control behaviorstates.
- $s_{co} \in S_{co}$ is the initial state.
- $E_{co}$ is a finite set of control behaviorevents.
- $A_{co}$ is a specific approaches of the antivirus behavior.
- The approach $A_{co}$ is denoted by $s_{co} \xrightarrow{e} s'_{co}$: $a_{co}$.

In statecharts, a specific approach is depicted by s[e]/a. State s $\in$ S presents a specific approach a $\in$ A in each behaviorby the event e $\in$ E.

The specific approaches on preventive and control behaviors are the same. The only deference between these approaches is connection between sates and transitions of each behavior.

**Definition 5** (mapping process): Let $P_{CO}$ be the set of paths by starting each state in the control behaviorof an antivirus system. The mapping process is defined by using the following function (1):

MP: $S_{co} \rightarrow 2^{Pdg \vee Pdi}$     (1)

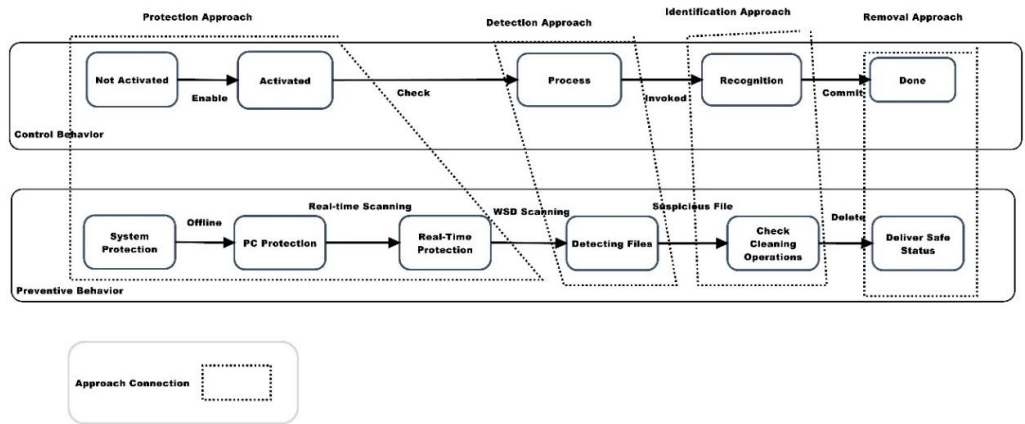

Figure 3. The example of mapped behavioral modeling in the antivirus system

The mapping function MP associates each state in the control behavior with a set of possible paths in the preventive behavior. In figure 3, we see a mapping method in the antivirus system approach where Not Activated and Activated states in the control behavior is associated with some states in the preventive behavior which are these: system protection → PC protection → real-time protection.

In figure 2, the states done and aborted are final state of control behavior (as a leaf node in the tree). For example, in figure 2 a path tree is: *not activated → activated → process → recognition → done*. In this path states have been terminated to done state at the end. This true path is a performance of the control behavior. There is a loop for states *process → recognition → process → recognition → process → recognition → process* …., which has not created a performance process of the control behavior.

In the antivirus system behavior we define four approaches. These approaches are Protection Approach, Detection Approach, Identification Approach and Removal Approach. Connections between preventive and control behaviors is possible by these approaches. Each of these approach are mapped on the states of preventive behavior and control behavior.

As we show in figure 3, the Protection Approach includes Not Activated and Activated states in control behavior and System Protection, PC Protection and Real-Time Protection in preventive behavior. Detection Approach includes Process state in control behavior and Detecting Files state in preventive behavior. Identification Approach includes Recognition state in control behavior and Check Cleaning Operations in preventive behavior and finally the removal Approach includes Done state in control behavior and Deliver Safe Status state in preventive behavior.

## 3. CONCLUSIONS AND FUTURE WORKS

In this paper, we present a behavioral model for the protection services in antivirus system which has two behavioral models, preventive behavior and control behavior. Then we tried to formal these models, and finally by explaining some definitions we illustrate the way these two behavioral models connected together. In future we try to verify these behavioral models by using NuSMV model checkers and Kripke structure techniques.

## REFERENCES

[1]   P. Szor, the Art of Computer Virus Research and Defence: Addison-Wesley Professional, 2005.

[2]   W. Wang, P.-t. Zhang, Y. Tan, and X.-g. He, "Animmune local concentration based virus detection approach," Journal of Zhejiang University SCIENCE C, vol. 12, pp. 443-454, 2011/06/01 2011.

[3]   X.-s. Zhang, T. Chen, J. Zheng, and H. Li, "Proactive worm propagation modelling and analysis in unstructured peer-to-peer networks," Journal of Zhejiang University SCIENCE C, vol. 11, pp. 119-129, 2010/02/01 2010.

[4]   T. Dube, R. Raines, G. Peterson, K. Bauer, M. Grimaila, and S. Rogers, "Malware target recognition via static heuristics," Computers & Security, vol. 31, pp. 137-147, 2// 2012.

[5]   J. J. C. H. Ryan, T. A. Mazzuchi, D. J. Ryan, J. Lopez de la Cruz, and R. Cooke, "Quantifying information security risks using expert judgment elicitation," Computers & Operations Research, vol. 39, pp. 774-784, 4// 2012.

[6]   F. B. Schneider, "Enforceable security policies," ACM Trans. Inf. Syst. Secur., vol. 3, pp. 30-50, 2000.

[7]   P. K. Singh and A. Lakhotia, "Analysis and detection of computer viruses and worms: an annotated bibliography," SIGPLAN Not., vol. 37, pp. 29-35, 2002.

[8]   E. Filiol, "Viruses and Malware," in Handbook of Information and Communication Security, P. Stavroulakis and M. Stamp, Eds., ed: Springer Berlin Heidelberg, 2010, pp. 747-769.

[9]   S. H. Sellke, N. B. Shroff, and S. Bagchi, "Modelling and Automated Containment of Worms," Dependable and Secure Computing, IEEE Transactions on, vol. 5, pp. 71-86, 2008.

[10]  C. Livadas and N. Lynch, "Formal verification of safety-critical hybrid systems," in Hybrid Systems: Computation and Control. vol. 1386, T. Henzinger and S. Sastry, Eds., ed: Springer Berlin Heidelberg, 1998, pp. 253-272.

[11]  Z. Qianchuan and B. H. Krogh, "Formal verification of Statecharts using finite-state model checkers," in American Control Conference, 2001. Proceedings of the 2001, 2001, pp. 313-318 vol.1.

[12]  S. Flores, S. Lucas, and A. Villanueva, "Formal Verification of Websites," Electronic Notes in Theoretical Computer Science, vol. 200, pp. 103-118, 5/23/ 2008.

[13]  J. Morales, P. Clarke, Y. Deng, and B. M. Golam Kibria, "Testing and evaluating virus detectors for handheld devices," Journal in Computer Virology, vol. 2, pp. 135-147, 2006/11/01 2006.

[14]  J. Andronick, B. Chetali, and C. Paulin-Mohring, "Formal Verification of Security Properties of Smart Card Embedded Source Code," in FM 2005: Formal Methods. vol. 3582, J. Fitzgerald, I. Hayes, and A. Tarlecki, Eds., ed: Springer Berlin Heidelberg, 2005, pp. 302-317.

[15]  C. L. Heitmeyer, M. M. Archer, E. I. Leonard, and J. D. McLean, "Applying Formal Methods to a Certifiably Secure Software System," Software Engineering, IEEE Transactions on, vol. 34, pp. 82-98, 2008.

[16]  W. L. Yeung, "Behaviour al modelling and verification of multi-agent systems for manufacturing control," Expert Systems with Applications, vol. 38, pp. 13555-13562, 10// 2011.

[17]  W. L. Yeung, "A formal and visual modelling approach to choreography based web services composition and conformance verification," Expert Systems with Applications, vol. 38, pp. 12772-12785, 9/15/ 2011.

[18]  J. Bentahar, H. Yahyaoui, M. Kova, and Z. Maamar, "Symbolic model checking composite Web services using operational and control behaviours," Expert Systems with Applications, vol. 40, pp. 508-522, 2/1/ 2013.

[19]  A. Souri and N. J. Navimipour, "Behaviour al Modelling and Formal Verification of Resource Discovery in Grid Computing," Expert Systems with Applications, 2013.

[20]  J. B. Shukla, G. Singh, P. Shukla, and A. Tripathi, "Modelling and analysis of the effects of antivirus software on an infected computer network," Applied Mathematics and Computation, vol. 227, pp. 11-18, 1/15/ 2014.

## Authors

**Monire Norouzi** received her B.Sc. in Computer Engineering at University College of Nabi Akram, Tabriz, Iran in 2011. Currently, she receives M.Sc. in Software Engineering from Shabestar Branch, Islamic Azad University in Iran. Her main research interests are Software Analysis, Wireless Network and Sensor Network, verification and validation of Software Systems.

International Journal in Foundations of Computer Science & Technology (IJFCST), Vol.4, No.3, May 2014**Saeed Parsa** received the B.S. degree in mathematics and computer Science from Sharif University of Technology in Iran, and the M.S. and Ph.D. degrees in computer science from the University of Salford at England. He is an associate professor of computer science at Iran University of Science and Technology (IUST). His research interests include software engineering, soft Computing and algorithms.

**Ali Mahjur** received his B.S. and M.S. degrees in computer engineering from Sharif University of Technology (SUT), Iran, in 1996 and 1998, respectively. He has been a Ph.D. student in computer engineering at SUT since then. His research interests include Computer Architecture, Operating Systems, Memory Management Systems, and Programming Languages.85